# Accommodation of the service offered by the network for networked control systems

I. Diouri, JP. Georges, E. Rondeau

CRAN UMR 7039

## 1. Introduction

Networked Controlled systems (NCSs) are composed of actuators, sensors and controllers distributed over the network. A NCS is a control system in which the regulation loops are closed by the communication network (Zhang et al., 2001). However, it is necessary to ensure the stability of the process control, taking into account the performances of the network (Branicky et al., 2003).
NCSs are more and more used in industrial applications because they present several advantages like to reduce the wiring costs (Gallara, 1984), to make easy the diagnosis and the maintenance of systems, and also to improve the modularity and the flexibility in the systems design.

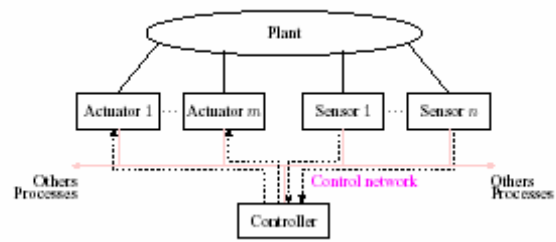

Figure 1: A typical NCS setup and information flows (Zhang et al, 2001)

The main problem of the NCSs is the network induced delays. These delays can degrade the performances of the systems; they can even lead to an unstable behaviour, especially for hard time-constrained processes. Thus, it is necessary to take into account these delays (Lian et al., 2001) in the design of the control law (Lian, 2001)(Cervin, 2003).

Network induced delays depend on:
- the communication protocol.
  NCSs can use any kind of networks which can either be determinism (traditional fielbuses like CAN, Profibus) or non-determinism (for instance, Ethernet). The latter one makes difficult the delay estimation.
- the traffic load.
  In addition to the medium access method, the load of the network influes on the delay. Indeed, traffic generated by unknown applications can disturb the performances offered to the real-time frames.

Ethernet is more and more used in the NCSs and then is studied in this paper. The CSMA/CD protocol used by the Ethernet (IEEE standard 802.3) (IEEE 2002) leads to a non-determinism access due to collisions. To avoid this problem, switched Ethernet network combined with full-duplex mode as defined in the IEEE standard 802.1D (IEEE 1998) is used.
Another interest of the Ethernet switch technology is to be able to implement mechanisms of Classification of Service (CoS) (IEEE 802.1p).

The CoS enables to manage different frames according to their priority. This management uses scheduling policies such as the Strict Priority or the Weighted Round Robin. In previous work – *deliverable D5.1* - (Georges et al., 2005b) proposed a method to calculate upper-bounded end-to-end delays for switched Ethernet networks with Strict Priority policy by using the Network Calculus. But the Strict Priority policy can lead to famine situations for the non real-time applications. The goal of this paper is to maximize the bandwidth allocation for unconstrained frames in guarantying that the control constraints are respected. The weighted round robin policy (WRR) (Demers et al., 1989) manages the network performances by adjusting the number of frames forwarded for each flow according to the frames priority. In this paper, we focus on switched Ethernet network implementing the Classification of Service (IEEE 802.1p) based on a Weighted Round Robin policy, as implemented by the Cisco Catalyst 2950 switches.

The Classification of Service in switched Ethernet networks is presented in the section 2. Then, the section 3 the WRR for which an evaluation and optimization method is proposed in the section 4. The section 5 discusses about the application of the method.

## 2. Classification of Service in Switched Ethernet networks

The native Ethernet does not implement any priority mechanism. Non standardized solutions have been proposed: adapting the interframe gap (smaller for high priority frames), modifying the *Binary Exponential Backoff* algorithm (the waiting time is not randomly calculated, but in relation with the priority), or using a variable length of the preamble (smaller for high priorities). Another approach consists in using a Time Division Multiple Access method over the native CSMA/CD protocol: pre-allocated time-slots are defined for the transmission of time-critical data.

Nevertheless, the evolution of Ethernet to segmented architectures and the definition of the *Virtual Local Area Networks* (VLAN) have led to the birth of a new standards set (802.1D/p, 802.1Q) in which new encapsulation fields are added to the classical frame (IEEE, 2003). One of these fields is specified in order to support 8 priority levels associated to 8 types of applications (voice, video, network management, best effort, etc.). The number of Classes of Service may be different to the number of priority levels, and also different for each port. That is why the standard also recommends a mapping between classes, priority and ports queues.

The next point is the scheduling policy that will be used to forward the frames at the output port regarding their dedicated priorities. (IEEE, 2003, section 8.6.6) defines two items.
- for a given supported value of traffic class, frames are selected from the corresponding queue for transmission only if all queues corresponding to numerically higher values of traffic class supported by the port are empty at the time of selection;
- for a given queue, the order of which frames are selected shall maintain the incoming ordering.

It means the scheduling policy defined is the *Strict Priority (SP)* algorithm and the policy must be FIFO for a given queue. But the standard enables to implement other algorithms. The main drawback of the SP algorithm is that it can lead to the impossibility for the lowest priority queues to be forwarded. To resolve it, CoS switches implement a supplementary policy: the *Weighted Fair Queuing (WFQ)*. In the Fair Queuing algorithms, the service offered to the high priority queues is moderated as following. A weight is associated to each queue. Then the scheduler gives to each queue (from the highest priority to the lowest) a bandwidth determined by its associated weight.

The Weighted Fair Queuing, initially proposed in (Demers et al., 1989), is also known as the Packetized Generalized Processor Sharing (PGPS). It is based on the conceptual algorithm called the Generalized Processor Sharing (GPS) (Parekh et al., 1993). However practical implementations of WFQ in today's switch products are based on a Weighted Round Robin which is more simple.

## 3. Weighted Round Robin policy

In a round robin policy, packets are pushed in queues according to their priority level. Then, the server pools the different queues according to a cyclic sequence (using a precomputed order defined by the queues priorities) in an attempt to serve one packet for each non empty queue. Even if this algorithm respects the fairness quality, no flexibility is integrated. Moreover, the fairness can be damaged with variable packet lengths. To improve the lack of flexibility of a simple round robin policy, the Weighted Round Robin (WRR) (Demers et al., 1989)(Katevenis et al., 1991) associates a weight $w_i$ on each flow $i$. Now, the WRR server will attempt to serve a flow $i$ with a rate of $\frac{w_i}{\sum_j w_j}$ before looking for the following queue. Comparing to PGPS, delays could be more important since if the system is heavily loaded and a frame just misses its slot, it will have to wait its next slot, i.e. a cycle.

In this paper, we will study a WFQ policy *based on a per-priority queuing and a weighted round robin scheduling*. This implementation is typical of switch products, like the *Cisco Catalyst 2950*. As shown in figure 2, the WRR gives a priority $i$ to each flow. It serves all the flows in a cyclic way, from the queue with the highest priority to the lowest. The number of frames that will be forwarded by the server for one queue $i$ is bounded by the number $\omega_i$. When the queue is empty, the scheduling protocol immediately processes the next queue.

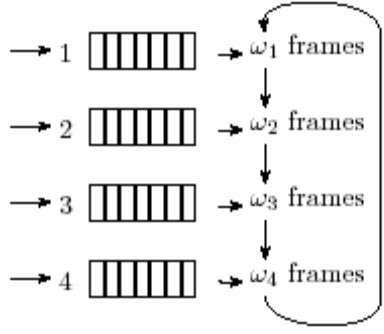

Figure 2: The WRR behaviour

Several extensions of the Round Robin policy have been proposed like for instance the Bit-by-bit Round Robin (BRR) (Demers et al., 1989) which tries to perform the RR algorithm bit by bit, but this is almost impossible to execute in a high speed network. Deficit Round Robin is proposed on the basis of RR, it uses a quantum concept to control the speed of packets travelling among flows. The Dynamic Weighted Round Robin (DWRR) (Li et al., 2005) is based on the flows generated by each station. To ensure the performances of this protocol, (Li et al., 2005) have given a method to determine the detecting interval; a very small interval could delay the system and a very long interval could cause a definitive un-balanced situation. However, since they are not yet implemented in switches, this study does not consider these approaches.

## 4. Service and delays offered by a WRR policy

In order to simplify the understanding of this section, only two flows, two queues and two priorities are taken into account. Anyway, the approach can be applied for more flows. In this study, the first flow corresponds to the traffic generated by the control of the distributed systems (real-time traffic). It will receive the highest priority and will be called flow 1. The second one, flow 2, gathers all the others traffic, i.e. background traffic. Its priority is lower than the control traffic.

The goal of this research is to satisfy the constraints of the control traffic without over estimating the service offered to this traffic. The main interest of this study is to avoid the specification of the background traffic which is difficult exactly identify. The proposed approach only needs to assume that $\omega_2$ background frames will be forwarded at each cycle and that the length of these frames corresponds to the Ethernet frames maximum length ($L$=1526 bytes). Let us define $\tau$ the processing time of one frame of the control flow and $\bar{\tau}$ the processing time of maximum length frame such that:

$$\tau = \frac{L}{C} \qquad \bar{\tau} = \frac{\bar{L}}{C}$$

The arrival of control frames is bounded by an affine arrival curve and the length of the frames $L$ is considered as constant. It means that the number of bytes arrived in the queue is upper-bounded by $b(t) = \sigma + \rho t$ where $\sigma$ is the maximum amount of traffic that can arrive in a burst and $\rho$ is an upper bound on the long-term average rate of the traffic.

Since the goal of this paper is to define a method to determine the weights $\omega_1$ and $\omega_2$ such as the time constraints of the distributed control are respected, this paper presents here an evaluation of the maximum delay for crossing a WRR server. This study is based on the Network Calculus theory introduced in (Le Boudec and Thiran, 2001).

*4.1 Minimal service curve*

According to the definition of Weighted Round Robin policy given in the section 3, the service offered to the control flow consists of sequence of waiting period (here called *vacation* period) and of *forwarding* period. The length of these periods mainly depends on the amount of frames in the queues and also on the length of the frames. The general pattern of the service curve is given in the figure 3.

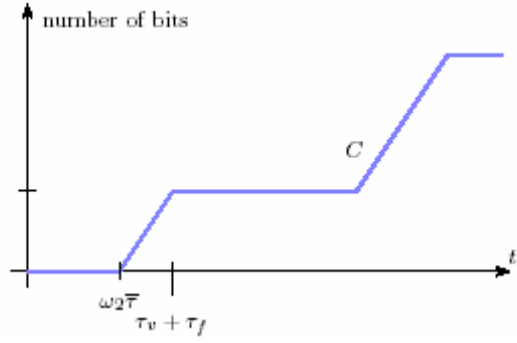
Figure 3: Service curves for the control flow

The service curve depends also on the vacation period $\tau_v$ and on the forwarding period $\tau_f$. The length of the sequence is given by $\tau_v + \tau_f$.

The main issue relies on the service offered by a WRR server. It depends on the arrival of control and background frames. This study requires identifying the minimal service offered to the flow 1, (the control traffic). This minimal service curve depends on the maximum vacation period and on the minimal forwarding period. The minimal forwarding period is related to the arrival of control data.

In this study, the arrival of control data is upper-bounded by the arrival curve $\alpha(t) = \sigma + \rho t$. The main characteristic of this assumption is that the maximum amount of traffic that can arrive in a burst is bounded by $\sigma$ and that the long-term average rate of the traffic $\rho$ is inferior to the output capacity $\rho < C$. This arrival curve is given in the figure 4.

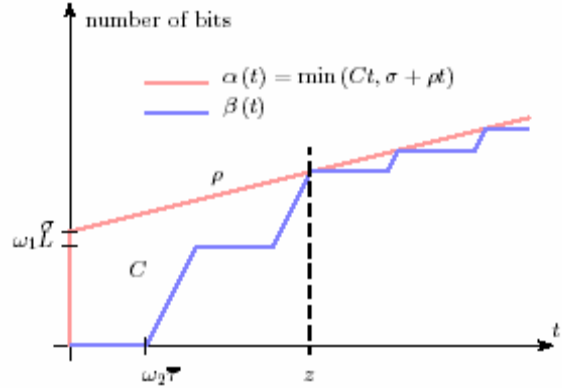
Figure 4: Arrival and service curves for the control flow

The figure 4 shows that two cases can be distinguished relatively to the amount of waiting data. Due to the initial burst, the WRR policy has to forward in a first time the maximum number of frames allowed per cycle since frames are waiting in the queue. This first case continues until that the queue is empty, i.e. that the service curve is equal to the arrival curve. This is important since otherwise it means that not enough resources have been allocated to the control flow and then the server is saturated. In a second time, the arrival of control frames follows the rate $\rho$. Considering that enough resources have been allocated and that $\rho < C$, the general form of the service curve is periodic. At the end of each sequence, the service curve reaches the arrival curve, which means that the control queue is empty.

In both cases, the general form of the service curve is defined by the following equation:

$$W_{\tau_v,\tau_f}(t) = \max\left( C\left(t - \tau_v \left\lceil \frac{t}{\tau_v + \tau_f} \right\rceil \right)^+, C\tau_f \left\lfloor \frac{t}{\tau_v + \tau_f} \right\rfloor \right) \tag{1}$$

In the figure 4, two processes are described: the *burst processing* and the *mean arrival processing*. Consider now the value of $\tau_v$ and $\tau_f$ in both cases.

*1) Burst processing*: According to the WRR definition, the vacation period corresponds to $\tau_v = \omega_2 \overline{\tau}$ since control frames are still waiting in the queue. It also means that the server will forward the maximum number of frames per cycle, which is defined by $\omega_1$. Hence, it gives:

$$\tau_v = \omega_2 \bar{\tau}, \quad \tau_f = \omega_1 \tau \tag{2}$$

and the service curve corresponding to the burst processing is given by: $\beta(t) = W_{\tau_v + \tau_f}(t)$.

As previously mentioned, this first processing time requires that the service curve reaches the arrival curve at a finite time, called z. This implies that the minimal value of $\omega_1$ is superior to:

$$k\omega_1 L \geq \sigma + \rho k \left(\omega_2 \bar{\tau} + \omega_1 \tau\right)$$

$$\omega_1 \geq \frac{\sigma + \rho k \omega_2 \bar{\tau}}{kL - k\rho\tau} \tag{3}$$

where $k$ corresponds to the number of cycle necessary to forward the burst and is defined by:

$$k = \left\lceil \frac{\sigma}{C\omega_1 \tau - \rho \omega_2 \bar{\tau}} \right\rceil$$

*2) Mean arrival processing*: We have now to consider the possibility that no control frames are waiting in the control queue at the end of the forwarding time of the background traffic. In this case, the WRR policy allows that another forwarding time of the background traffic can immediately start. Therefore the maximum vacation period $\tau_v$ depends on the maximum interarrival time of control frames, such that we have:

$$\tau_v = \left\lfloor \frac{L/\rho - \tau}{\omega_2 \bar{\tau}} \right\rfloor \omega_2 \bar{\tau} \tag{4}$$

Consider now the minimum forwarding period of the control flow. As shown in the figure 4, it is necessary that the service curve reaches the arrival curves in order to guarantee the service required by the control application. Therefore, $\tau_f$ consists here to the time necessary to forward the traffic received during the vacation period, such that:

$$\tau_f = \frac{\rho}{C - \rho \tau_v} \tag{5}$$

and the service curve corresponding to the mean arrival processing is given by:

$$\beta(t) = W_{\left\lfloor \frac{L/\rho - \tau}{\omega_2 \bar{\tau}} \right\rfloor \omega_2 \bar{\tau}, \frac{\rho \left\lfloor \frac{L/\rho - \tau}{\omega_2 \bar{\tau}} \right\rfloor \omega_2 \bar{\tau}}{C - \rho}}(t)$$

To note that the previous assumptions require that $\tau_f \leq \frac{\omega_1 L}{C}$, and also:

$$\frac{\omega_1 L}{C} \leq \frac{\rho \tau_v}{C - \rho}$$

$$\omega_1 \leq \frac{C \rho \tau_v}{L(C - \rho)} \tag{6}$$

*4.2 Maximum delay*

(Le Boudec et Thiran, 2001) noted that the delay that would be experienced by a bit arriving at time *t* if all bits received before it are served before it, is bounded by the horizontal deviation between the arrival curve and the service curve. The delay experienced by a frame corresponds to:

$$d(t) = \inf \{\Delta \geq 0 : \alpha(t) \leq \beta(t + \Delta)\}$$

The delay is then upper-bounded by $\overline{D} \leq \sup\{t \geq 0 : d(t)\}$. It will be assumed that the arrival of control of data is simply constraint by the arrival curve $\alpha(t) = \sigma + \rho t$. The figure 5 gives an overview of the horizontal deviation between the arrival curve and the service curve identified in the previous section.

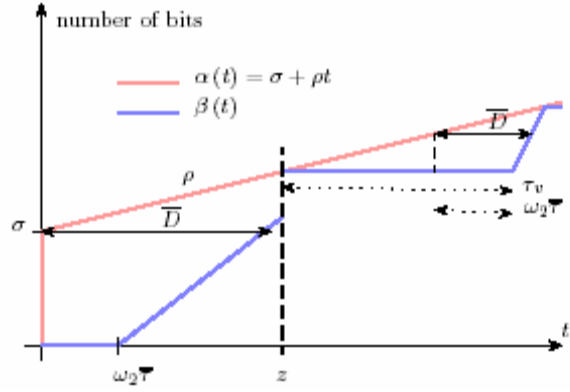

Figure 5: Delays for the control flow

*1) Burst processing*: It is noticed that a rate-latency service curve can be directly obtained from the equation (2). As shown in the figure 5, the service curve that is now considered is:

$$\beta_{R,T}(t) = \frac{\omega_1 L}{\tau_v + \tau_f}(t - \tau_v)^+$$

with $\tau_v$ and $\tau_f$ defined in the equation (2).

In (Georges et al., 2005a), we have shown that the delay for a given arrival curve $\alpha(t) = \sigma + \rho t$ and a given service curve $\beta_{R,T}(t) = R(t-T)^+$ is bounded by:

$$\overline{D} = (T - \tau_i) + \frac{\sigma + \rho_i \tau_i}{R} \quad (7)$$

The upper-bound given in the equation (7) is applied in this case with $R = \omega_1 L / (\omega_1 \tau + \omega_2 \overline{\tau})$ and $T = \omega_2 \overline{\tau}$. Hence the delay is upper-bounded by:

$$\overline{D} = \omega_2 \overline{\tau} + \frac{\sigma}{C} + \sigma \frac{\omega_2 \overline{\tau}}{\omega_1 L}$$

$$\overline{D} = \omega_2 \overline{\tau} + \frac{\sigma}{C} \frac{\omega_1 L + \omega_2 \overline{L}}{\omega_1 L} \quad (8)$$

The equation (8) shows that the maximum delay consists on two parts: the vacation period and the time to process the burst. A high value of $\omega_1$ will also contributes to reduce the time to forward the burst and hence the maximum delay, but in the same time, the equation (8) highlights that increasing the length of $\tau_f$ limits the bandwidth for the non real-time traffic.

*2) Mean arrival processing*: Using the service curve determined in the previous section, the delay is defined by:

$$d(t) = \inf\{\Delta \geq 0 : \rho_i t \leq C(t + \Delta - \tau_v)^+\}$$

Using the maximum operator $\vee$ ($a \vee b = \max(a,b)$), $\overline{D}$ can be decomposed in:

$$\overline{D} = d(0) \vee \sup\{0 < t \leq \tau_v : d(t)\} \vee \sup\{\tau_v < t \leq \tau_v + \tau_f : d(t)\}$$

$$= 0 \vee \sup\left\{0 < t \leq \tau_v : \inf\left\{\Delta \geq 0 : \Delta \leq \frac{(\rho - C)t}{C} + \tau_v\right\}\right\}$$

$$\vee \sup\left\{\tau_v < t \leq \tau_v + \tau_f : \inf\left\{\Delta \geq 0 : \Delta \leq \frac{(\rho - C)t}{C} + \tau_v\right\}\right\}$$

Since $\rho < C$ and according to the equation (4), it gives:

$$\overline{D} = 0 \vee \tau_v \vee \frac{\rho \tau_v}{C} = \tau_v = \left\lfloor \frac{L/\rho - \overline{\tau}}{\omega_2 \overline{\tau}} \right\rfloor \omega_2 \overline{\tau} \quad (9)$$

However, even if the result given in equation (9) is an upper-bound of the delay, it might be too much pessimistic. Indeed, the delay must be associated to a frame. In this second phase, since $\rho < C$, the delay is simply bounded by the forwarding time of one frame $\tau$ plus the time necessary to forward one time the background queue and not the whole vacation period. The formula of the maximum delay should not include time period for which no frames have been yet received. The horizontal deviation has to be only considered for frame as shown in the figure 5. Hence, the upper-bound which must be considered is:

$$\overline{D} = \omega_2 \overline{\tau} + \tau = \omega_2 \overline{\tau} + \frac{L}{C} \quad (10)$$

The equation (10) shows that the maximum delay consists of the vacation period plus the forwarding time of one control frame. The overall delay is also bounded by:

$$\overline{D} = \omega_2 \overline{\tau} + \tau = \left(\omega_2 \overline{\tau} + \frac{L}{C}\right) \vee \left(\omega_2 \overline{\tau} + \frac{\sigma}{C} \frac{\omega_1 L + \omega_2 \overline{L}}{\omega_1 L}\right) \quad (11)$$

*4.3 Departure curve*

Initially, only the arrival curve of the control flow is only known at the input of the network. On a multi-hop path (i.e. a path with several WRR servers), the arrival curve of the control traffic must be identified at the output of each WRR server. For that, (Le Boudec et Thiran, 2001) shows the output flow of a system is constrained by a departure curve $\alpha^*(t)$ (which is in fact the arrival curve of the following system), defined by:

$$\alpha^*(t) = \alpha(t) \oslash \beta(t) = \sup\{v \geq 0 : \alpha(t+v) - \beta(v)\}$$

The general form of the service curve used previously to identify the maximum delay is the rate-latency curve $\beta_{R,T}(t) = R(t - T)^+$. In (Georges et al., 2005a), we have shown that the departure curve for a given arrival curve $\alpha(t) = \sigma + \rho t$ and a given service curve $\beta_{R,T}(t)$ is bounded by:

$$\alpha^*(t) = \sigma + \rho T + \rho t$$

The previous equation shows that the crossing of a WRR server by an affine arrival curve leads to increase the maximum amount of data that can arrive in a burst relatively to the vacation period $\tau_v$. Moreover, according to the WRR algorithm, the number of control frames forwarded during a cycle is bounded by $\omega_1$. Therefore since in both cases we have $T = \omega_2 \overline{\tau}$, the control traffic is constrained at the output of a WRR server by:

$$\alpha^*(t) = \min\left(\omega_1 L + \rho t, \sigma + \rho \omega_2 \overline{\tau} + \rho t\right) \quad (12)$$

In this section, an upper-bound of the maximum delay, constraints on $\omega_1$ and a definition of the output burst have been provided. These results are applied in the next section in order to determine the value of $\omega_1$ which consumes the minimal effort for a given case study.

## 5. Illustration

In order to illustrate the computations introduced in section 4, the proposed method to fix the weights $\omega_1$ and $\omega_2$ on each WRR compliant devices is applied to the architecture shown in the figure 6.

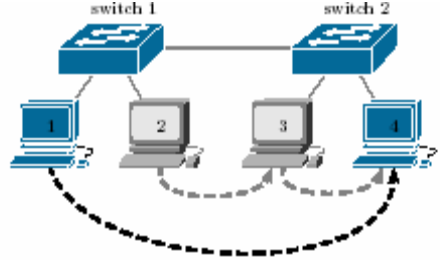

Figure 6: The case study

In the figure 6, four workstations are interconnected on a switched Ethernet network. Three flows are considered: one real-time between stations 1 and 4, and two background traffics (between stations 2 and 3 and between stations 3 and 4). It is assumed here that the switches implement the WRR policy on each port like the Cisco Catalyst 2950 switches. The links are configured at C=10 Mb/s in the full-duplex mode. The station 1 sends one frame of L=72 bytes each T=5 ms. The goal is to determine the weights $\omega_1$ and $\omega_2$ on each switches such that the end-to-end delay for the real-time traffic is inferior to T=5 ms and such that the service offered by the switches to the background traffics is the highest as possible.

The first step consists to identify the arrival curve of the real-time traffic on the first switch. As shown by (Le Boudec et Thiran, 2001), an affine arrival curve might be derivate from a periodic arrival ($\alpha(t) = Lv_T(t) = L\left\lfloor\frac{t}{T}\right\rfloor$ where $T$ corresponds to the period and $L$ to the frames length). Using the relation $\alpha(t) = L + L\left\lfloor\frac{t}{T}\right\rfloor$, it gives $\alpha(t) = \sigma + \rho t$ with $\sigma = L$ and $\rho = 115.2\ kb/s$. For the arrival curve $\alpha^*(t)$ on the second switch, the equation (12) and the relation $\sigma = L$ give directly $\alpha^*(t) = \omega_1 L + \rho t$ where $\omega_1$ represents the maximum number of time-constrained frames forwarded during one cycle for the first switch.

Then an iterative method was applied as a first result to determine the weights $\omega_1$ and $\omega_2$ on each switch. The idea consists to fix an initial value of $\omega_2$, to determine the corresponding $\omega_1$ using the relation given in the equations (3) and (6) and to compute the delays to cross each switch. We finally obtain that the WRR policy has to be configured with $(\omega_1 = 2,\ \omega_2 = 1)$ for the first switch and with $(\omega_1 = 9,\ \omega_2 = 2)$ for the second one. These result stands for a maximum delay for real-time frames of 1.8888 ms to cross the first switch and of 3.099 ms to cross the second one.

As a result, the method proposed in this paper ensures that the end-to-end delay is inferior to the bound defined by the real-time application. Moreover, it is interesting to note that the bandwidth offered by the switches to the background traffics is equal to 9.138 Mb/s for the first one and to 8.249 Mb/s for the second one, such that the overall bandwidth corresponds to 8.249 Mb/s. It shows that the conservative values of let anyway the large amount of bandwidth to the other flows.

## 6. Conclusion

The major result of this paper is a method to determine the weights of a Weighted Round Robin policy such that the time-constraints of the applications are guaranteed and such that the bandwidth offered to the other traffics is maximized. The study focus on switched Ethernet network implementing the Classification of Service (IEEE 802.1p) based on a Weighted Round Robin policy. Real experimentations are in progress.

In the context of networked control systems, the next step will be to define an algorithm enabling to automatically associate a time constraint requirement with a priority level and a weighted value. More advanced optimization algorithms will be also study for the determination of the weights.